\documentclass[aps,prx,twocolumn]{revtex4-1} 
\usepackage{graphicx}
\usepackage{amsmath,amsfonts,amssymb}
\usepackage{bm}
\usepackage{color}

\usepackage{xcolor}

\usepackage[normalem]{ulem}

\usepackage{soul}

\begin{document}

\newcommand{\matteo}[1]{{\color{green} #1}}
\newcommand{\commento}[1]{{\color{red} #1}}


\title{Hidden velocity ordering in dense suspensions of self-propelled disks}

\author{Lorenzo Caprini}
\email{lorenzo.caprini@gssi.it}
\affiliation{Universit\'a di Camerino, Dipartimento di Fisica, Via Madonna delle Carceri, I-62032 Camerino, Italy}
\author{Claudio Maggi}
\affiliation{CNR-Istituto Sistemi Complessi, P.le A. Moro, I-00185, Rome, Italy}
\author{Umberto Marini Bettolo Marconi}
\affiliation{Universit\'a di Camerino, Dipartimento di Fisica, Via Madonna delle Carceri, I-62032 Camerino, Italy}
\author{Matteo Paoluzzi}
\affiliation{CNR-Istituto Sistemi Complessi, P.le A. Moro, I-00185, Rome, Italy}
\author{Andrea Puglisi}
\affiliation{CNR-Istituto Sistemi Complessi, P.le A. Moro, I-00185, Rome, Italy}
\date{\today}


\begin{abstract}
Recent studies of the phase diagram for spherical, purely repulsive,
active particles established the existence of a transition from a
liquid-like to a solid-like phase analogous to the one observed in
colloidal systems at thermal equilibrium, in particular in two
dimensions an intermediate hexatic phase is observed.  Here, we present
evidence that the active dense phases (solid, hexatic and liquid) exhibit
interesting dynamical anomalies. 
First, we unveil the growth - with density and activity - of ordered domains where the particles'
velocities align in parallel or vortex-like patches. Second, when activity
is strong, the spatial distribution of kinetic energy becomes
heterogeneous with high energy regions correlated to defects of
the crystalline structure.  This spatial heterogeneity is accompanied
by temporal intermittency, with sudden peaks in the time-series of 
kinetic energy. The observed dynamical anomalies are not present
in a dense equilibrium system and cannot be detected by considering
only the structural properties of the system.
\end{abstract}
\maketitle

\section{Introduction}
The dynamics of  
colloidal particles at high densities has been
largely explored, theoretically, numerically and experimentally, in
the last decades~\cite{lowen1994melting,
  gasser2009crystallization}. At thermodynamic equilibrium, the
Mermin-Wagner theorem rules out the existence of a crystalline phase
in two dimensional systems, characterized by long-range translational
order~\cite{mermin1966absence,mermin1968crystalline}. 
As shown by Halperin, Nelson~\cite{PhysRevLett.41.121}, and Young~\cite{PhysRevB.19.1855}, in two dimensions the melting transition proceeds
 via two continuous Berezinskii-Kosterlitz-Thouless~\cite{berezinsky1970destruction,kosterlitz1973ordering}  transitions driven by topological defects, i.e., 
a hexatic-liquid transition, with a quasi-long range orientational order, and a solid-hexatic transition, characterized by quasi-long range translational order and long-range 
orientational order~\cite{PhysRevB.19.1855,gasser2010melting}.
Density and temperature are the control parameters to move from liquid
to hexatic and to solid-like aggregation phases.  
This scenario has been verified employing dense suspensions of equilibrium colloids~\cite{zahn1999two, dullens2017two}.

Recently, the study of two dimensional systems of self-propelled
particles at high packing fractions has attracted the attention of the
active matter community~\cite{marchetti2013hydrodynamics, gompper20202020}, since it may offer interesting engineering
applications, for instance, in the design of new materials~\cite{bechinger2016active}.  
Although most of the
experimental studies have so far focused on the low densities regime, some novel experiments have investigated Janus particles also in the case of very dense suspensions~\cite{klongvessa2019relaxation}. 
Another interesting
class of high-density non-equilibrium systems is represented by driven
granular media where mono-disperse polar grains under shaking~\cite{deseigne2010collective} display persistent motion. 
It is also worth to mention some recent studies of artificial microswimmers at such densities~\cite{briand2016crystallization}, revealing an intriguing experimental scenario for non-equilibrium aggregation. 
Specimens of active matter systems at very high densities are
very interesting also in the biological realm.  Typical examples are
tissues composed of highly packed eukaryotic cells.  Particular attention has been devoted to the dynamics of
confluent monolayers~\cite{angelini2011glass, garcia2015physics},
which slows down as density increases.  
More recently, an
amorphous ``solidification'' process has been investigated during the
process of vertebrate body axis elongation~\cite{mongera2018fluid},
where cells become solid-like.

\begin{figure*}[!t]
\centering
\includegraphics[width=0.98\linewidth,keepaspectratio]{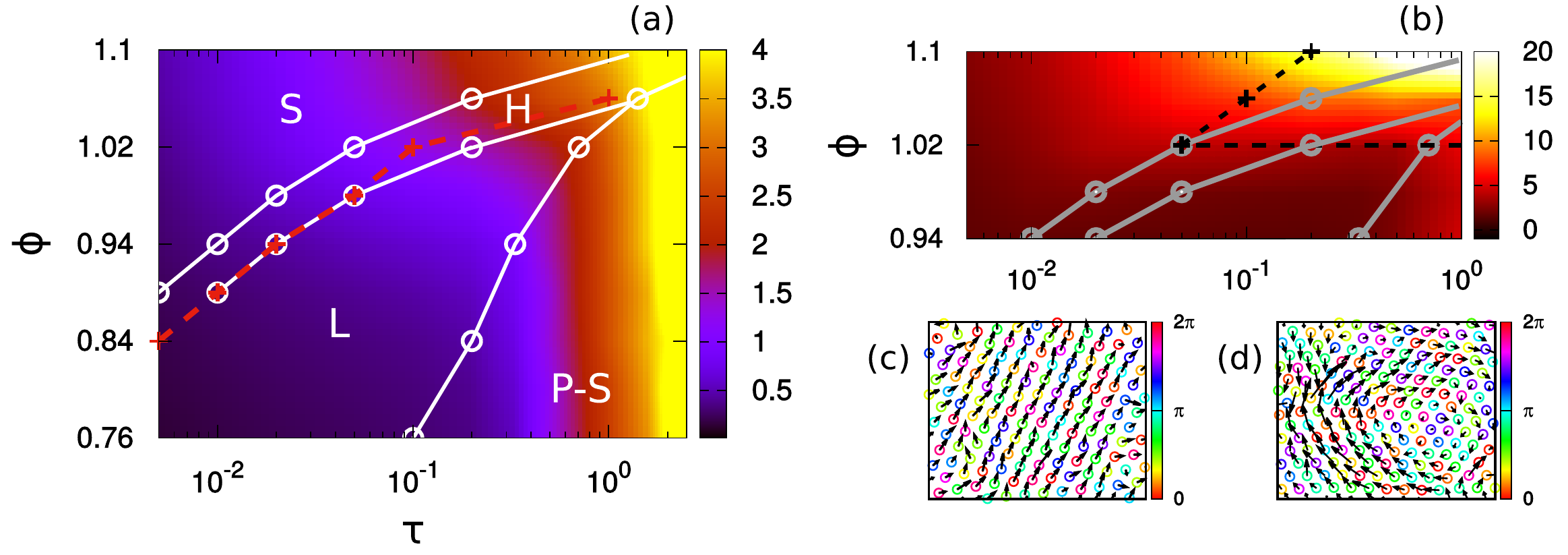}
\caption{ 
  Dynamical phase diagram. In panel (a), we
  plot the phase diagram as a function of $\tau$ and $\phi$, coloring
  each region depending on the size of the aligned domains, quantified
  by the parameter $R/\sigma$. The symbols S, H, L, and P-S denote solid,
  hexatic, liquid and phase-separated phases, respectively, and are
  separated by solid lines. The dashed red line is obtained monitoring
  the pair correlation function, $g(r)$.  
  Panel (b) focuses on the top
  region of the phase diagram and is colored depending on the
  correlation length, $\ell$, of the spatial velocity correlation. The
  dashed black line delimits the regions where kinetic energies show
  an intermittent phenomenology.  Panels (c) and (d) show two zooms of
  the snapshot configuration realized at $\phi=1.1$ and $\tau=1$.
  Colors encode the orientation of the self-propulsion while the black
  arrows represent the velocity vectors.  Panel (c) shows an aligned
  domain, while panel (d) a vortex-like structure in the pattern of
  the particles' velocities.  Simulations are obtained with $v_0=50$
  and the inter-particles interaction discussed in the main text.  }
\label{fig:velocitysnapshots_extreme}
\end{figure*}

Theoretical approaches in the statistical physics of active matter focus
on simplified models of self-propelled particles, the Active Brownian
Particle (ABP) being one of the most studied. Despite the existence of
a vast literature concerning the regime of moderate packing fractions,
ABP dynamics in the high-density regime is by far less
explored. 
For instance, active crystallization is studied in Ref.~\cite{bialke2012crystallization}, where a
a shift of the liquid-solid transition line towards larger densities with respect to
the Brownian counterpart is revealed.
Besides, this transition is accompanied by a
true non-equilibrium phenomenon: liquid and solid phases are separated
by a region where the suspension is globally ordered but bubbles of
``liquid'' still persist.  A more detailed analysis revealed the
occurrence of traveling crystals~\cite{menzel2013traveling,
  menzel2014active}, accompanied by the transition to rhombic, square
and even lamellar patterns.  This phenomenology has been recently
confirmed by experiments realized with vibrating granular disks~\cite{briand2018spontaneously}.

The construction of the phase diagram of the ABP model~\cite{digregorio2018full, cugliandolo2017phase, stenhammar2014phase},
follows the idea that its behavior~\cite{klamser2018thermodynamic}
for small active forces resembles that of passive Brownian particles
with the occurrence of ``gas-like'', ``liquid-like", ``hexatic-like''
and ``solid-like'' phases, with a shift of the transition lines
towards larger densities when activity increases~\cite{digregorio2018full, klamser2018thermodynamic}.  On the contrary,
for large self-propulsion (but moderate densities) an unexpected
phenomenon occurs:
the system phase separates even in the absence
of attractive interactions~\cite{redner2013structure,buttinoni2013dynamical, palacci2013living,bialke2015active, ginot2018aggregation, siebert2017phase, cugliandolo2017phase,speck2016collective,chiarantoni2020work}
(the so-called Motility Induced Phase Separation
(MIPS)~\cite{fily2012athermal, cates2015motility, gonnella2015motility, ma2020dynamic}).  Differently, at high densities but far from
equilibrium, ``standard'' crystallization seems to occur~\cite{bialke2012crystallization, digregorio2018full}.  The general
picture suggested by these studies is that
the high-density equilibrium scenario extends, qualitatively identical, to active systems, with the only difference that self-propulsion may destabilize the ordered phases or induce a phase-separation.

Some other studies suggest that active dense
phases in the MIPS region display a richer picture with respect to passive phases in the coexistence region. For instance, a quite different behavior can be found by analyzing the pressure~\cite{solon2015pressure, solon2015pressure_PRL} and
the interfacial tension between the gas and the cluster phase~\cite{bialke2015negative, patch2018curvature}.
Recently, the study of
the particles' velocities has revealed unexpected features which are
certainly absent in equilibrium fluids, e.g.  the different kinetic
temperatures inside and outside a cluster~\cite{mandal2019motility},
and the spontaneous alignment of velocities in the phase-separated
regime~\cite{caprini2019spontaneous}.

\subsection{Summary of results}

Our main findings can be summarised by the introduction of dynamical 
information about the particles' velocities
into the structural phase diagram, as shown
in Fig.~\ref{fig:velocitysnapshots_extreme}. The phase diagram
concerns the two-dimensional high-density regimes (both homogeneous
and phase-separated), with two main control parameters: the
persistence time, $\tau$, of the active force (which is proportional to the
P\'eclet number and inversely proportional to the rotational
diffusivity) and the packing fraction, $\phi$. We recall the definition
of $\phi=N \sigma^2\pi /4L^2$, being $N/L^2$ the number density and
$\sigma$ the particles' diameter.
Our ``augmented'' phase diagram
challenges the widespread idea that the structural properties alone are
enough to describe the dense phases of self-propelled particles,  
and suggests that a richer picture 
is obtained by including velocity correlations which, in turn, represent an exquisitely off-equilibrium feature of active systems.

Panel (a) of Fig.~\ref{fig:velocitysnapshots_extreme} portrays the
phase diagram as a function $\phi$ and $\tau$, which
reproduces~\cite{digregorio2018full}, with three homogeneous phases,
i.e. the solid phase (S), the hexatic phase (H) and the liquid phase (L), and a non-homogeneous regime with MIPS-like phase coexistence
(P-S), see Sec.~\ref{Sec:InteractingABPparticles}~B for details.  Our first finding is that the
alignment of particles' velocities discovered
in~\cite{caprini2019spontaneous} in the P-S regime is observed also in
the \emph{homogeneous}  liquid, hexatic and solid 
phases.  Particles' velocities form patterns arranging in aligned
or vortex-like domains, as shown by the arrows in panels (c) and (d)
of Fig.~\ref{fig:velocitysnapshots_extreme}, even if the orientation
of self-propulsion has negligible order (see color coding in the same
panels).  The size of aligned domains - quantified by $R$, defined
later, encoded by colors in panel (a) - grows as $\tau$ and $\phi$
increase, 
as discussed in detail in
Sec.~\ref{Sec:highdensities_numerics}~A.  
In the homogeneous
liquid configurations, the size of the aligned velocity domains is
rather small as a result of the absence of translational order, at
variance with the solid (denser) case where the sizes reach large
values.  In the non-homogeneous configurations, only the growth of
$\tau$ induces the increase of the correlation length as a result of
density saturation. We recall that order in the velocity field is
absent in any passive Brownian suspension, even at high densities: it
is, in fact, a pure non-equilibrium feature due to the presence of propulsion forces in the active dynamics. 
Interestingly, a similar effect is also observed in fluidized granular
materials, but it is caused by the presence of dissipative interactions~\cite{puglisi12,manacordabook,plati2019dynamical}.
Fig.~\ref{fig:velocitysnapshots_extreme}~(b) focuses on the top part
(largest densities) of the phase diagram: the color encodes
different information here, i.e.  the correlation length $\ell$ of the
spatial velocity correlation which takes into account also kinetic
energy (square modulus of velocity) and not only the orientation of the
velocity vectors as in the case of $R$ in panel (a).  We observe that
$\ell$ increases in the solid phase and saturates in the hexatic or
liquid phases, because of the absence of both translational and/or
orientational order (it increases again in the phase-separated regime as explained in detail in Sec.~\ref{Sec:highdensities_numerics}~C). In the same panel, a
dashed black line delimits a region where heterogeneous spatial
distributions and temporal intermittent behaviors of the kinetic energy
are observed: interestingly, these anomalies in the kinetic energy field
are correlated to the structural defects of the crystal arrangement (see details in Sec.~\ref{Sec:highdensities_numerics}~B).

The article is structured as follows: in
Sec.~\ref{Sec:InteractingABPparticles}, we introduce the ABP model for
interacting self-propelled particles, summarizing the structural
properties of the system.  In Sec.~\ref{Sec:highdensities_numerics},
we present a detailed study of all the dynamical anomalies in the
velocity orientation, velocity vector and kinetic energy fields,
correlating these anomalies to the different structural properties of
the system. A theoretical approach is also presented in
Sec.~\ref{Sec:highdensities_numerics}~C 
that allows us to predict the
features of the spatial correlation functions of the velocity field. 
Section~\ref{sec:conclusion} is devoted to conclusions and perspectives.

\section{The system of interacting self-propelled particles}\label{Sec:InteractingABPparticles}

We study a system of $N$ interacting ABP disks in two dimensions
moving in a fluid at high viscosity (low-Reynolds number).  We neglect
both hydrodynamic interactions among the particles and inertial terms~\cite{bechinger2016active}.
The center of mass of each disk, $\mathbf{x}_i$, evolves according to
the following stochastic differential equation:
\begin{equation}
\label{eq:x_dynamics_ABP}
\gamma\dot{\mathbf{x}}_i = \mathbf{F}_i + \mathbf{f}^a_i \,,
\end{equation}
where $\gamma$ is the drag coefficient of the fluid and the effect of
the thermal noise due to the solvent is assumed to be much smaller
than the effect due to the random active force $ \mathbf{f}^a_i$~\cite{bechinger2016active}. The
term $\mathbf{F}_i$ represents the force contribution due to steric
interactions, such that, $\mathbf{F}_i=-\nabla_i U_{tot}$, being
$U_{tot}= \sum_{i<j} U(|{\mathbf x}_{ij}|)$, with ${\mathbf x}_{ij}=
\mathbf{x}_i -\mathbf{x}_j$.
 
Following several studies in the literature~\cite{redner2013structure, caprini2019spontaneous}, we choose $U(r)$ as a truncated and shifted Lennard-Jones potential:
\begin{equation}
\label{eq:potentialshape}
U(r)=
\begin{cases}
&4\epsilon\left[\left(\dfrac{\sigma}{r}\right)^{12}-
 \left(\dfrac{\sigma}{r}\right)^{6}\right] + \epsilon,  \quad \,\, r\leq2^{1/6}\sigma\\
&0   \,,     \qquad\qquad\qquad\qquad \qquad\quad  r\geq 2^{1/6}\sigma \, .
\end{cases}
\end{equation}
The constant $\sigma$ represents the nominal particle diameter while
$\epsilon$ is the typical energy scale of the interaction.  For numerical
convenience, both these parameters are set to one in the simulations.
Even in very packed configurations, each particle can only interact
with its first neighbors due to the truncated potential.

The self-propulsion force, $\mathbf{f}^a_i=\gamma v_0 \mathbf{n}_i$,
evolves with ABP dynamics: $v_0$ is
the modulus of the speed
induced by the self-propulsion for a force-free particle,
$\mathbf{n}_i$ is a unit vector of components $(\cos{\theta_i},
\sin{\theta_i})$. 
The orientational angle, $\theta_i$, performs
angular diffusive motion described by:
\begin{equation}
\label{eq:theta_dynamics}
\dot{\theta}_i= \sqrt{2D_r} \,\xi_i  \,,
\end{equation}
where $\xi_i$ is a white noise with unit variance and zero average.
The constant $D_r$ represents the rotational diffusion coefficient
and its inverse defines the typical 
relaxation time, $\tau=1/D_r$, of the active force~\cite{farage2015effective}. 

We remark that no explicit aligning force is included in the present
model at variance with Vicsek-like models where particles' velocities
are forced to align to the mean orientation of surrounding particles'
velocities~\cite{vicsek2012collective, gregoire2004onset}.  Thus, in
contrast with Vicsek-like models, the dynamics~\eqref{eq:x_dynamics_ABP} does not produce any polarization of the
directors $\mathbf{n}_i$.
We also avoid employing any form of self-alignment between the
particle velocity and the self-propulsion force responsible for
orientation-velocity ordering as recently proposed in~\cite{lam2015self, giavazzi2018flocking}.

\subsection{The effective velocity dynamics of the particle }

In order to obtain theoretical predictions and interpret the
results, following~\cite{caprini2019spontaneous}, it is useful to
switch from the set of variables $\{\mathbf{x}_i, \mathbf{f}^a_i\}$ to
the transformed variables $\{\mathbf{x}_i, \mathbf{v}_i\}$ eliminating
the self-propulsions in favor of the particles' velocities,
$\mathbf{v}_i=\dot{\mathbf{x}}_i$. We underline that the vectors $\mathbf{f}^a_i$ and
$\mathbf{v}_i$ are not aligned, because of the
interaction force $\mathbf{F}_i$. This is true, in particular, at high
densities.  The transformed dynamics reads:
\begin{flalign}
\label{eq:xv_dynamics_MIPS_varx}
\dot{\mathbf{x}}_i &= \mathbf{v}_i \\
\label{eq:xv_dynamics_MIPS}
\tau\gamma\dot{\mathbf{v}}_i &= - \gamma\sum_{j=1}^N{\boldsymbol{\Gamma}}_{ij}({\mathbf r}_{ij}) \mathbf{v}_j + \mathbf{F}_i +  \tau\gamma\mathbf{k}_i 
\end{flalign}  
where both $\mathbf{v}_i$ and $\mathbf{x}_i$ belong to the plane $xy$ and $\mathbf{r}_{ij}=\mathbf{x}_i-\mathbf{x}_j$. Each term $\boldsymbol{\Gamma}_{ij}$ is a $2\times 2$  matrix, whose elements are:
\begin{equation}
\label{eq:Gammadefinition_MIPS}
\Gamma_{ij}^{\alpha \beta}({\mathbf r}_{ij}) = \delta_{ij}\delta_{\alpha \beta} + \frac{\tau}{ \gamma} \nabla_{i\alpha} \nabla_{j \beta} U(|{\mathbf r}_{ij}|) \,,
\end{equation}
where Latin indices identifying the particles assume values $i,j=1,N$ 
while the Greek indices stand for spatial components $(x,y)$.
 The last addend in Eq.~\eqref{eq:xv_dynamics_MIPS}, $\mathbf{k}_i$, is a noise term which reads:
\begin{equation}
\mathbf{k}_i=v_0 \sqrt{2  /\tau} \,\boldsymbol{\xi}_i \times {\mathbf n}_i = v_0 \sqrt{2 /\tau} \,\boldsymbol{\xi}_i \times \frac{\gamma\mathbf{v}_i-\mathbf{F }_i}{\gamma v_0} \,.
\end{equation}
$\boldsymbol{\xi}_i$ is the stochastic vector with components $(0,0, \xi_i)$ normal to plane of motion.
At variance with the dynamics of $\mathbf{f}^a_i$, the modulus of $\mathbf{v}_i$ is not constant because of the term $\propto\boldsymbol{\xi}_i\times\mathbf{F}_i$.
The result can be easily generalized to the case of finite thermal noise, as shown in Appendix~\ref{appendix:changeofvariablesT}, using the strategy of~\cite{caprini2018active}.

The deterministic part of Eq.~\eqref{eq:xv_dynamics_MIPS}
represents the dynamics of an underdamped passive particle under the
action of a space-dependent friction.  Indeed, the first term in the
right-hand side of Eq.~\eqref{eq:xv_dynamics_MIPS} can be split into
a contribution $-\gamma \Gamma_{ii} \mathbf{v}_i$, representing a
generalized Stokes force acting on the $i$-th particle plus a second contribution,
$-\gamma \sum_{j\neq i}^N{\Gamma}_{ij}({\mathbf r}_{ij})
\mathbf{v}_j$, which depends on particles' relative positions and velocities
and vanishes in a passive system.

The ABP equation~\eqref{eq:xv_dynamics_MIPS} for the velocity strongly
resembles the analogous equation of another schematic model of
self-propelled particles, namely the Active Ornstein-Uhlenbeck model
(AOUP)~\cite{marconi2015towards,marconi2017heat,marconi2016velocity, fodor2016far, wittmann2019pressure,
  berthier2019glassy, caprini2019entropy, woillez2019nonlocal, caprini2019activechiral,maggi2017memory}. Such a connection has recently been established in
some studies~\cite{caprini2019comparative, das2018confined}: the only
difference between ABP and AOUP arises from the noise term
$\mathbf{k}_i$.  In the latter case, $\mathbf{k}_i$ is a simple white
noise acting on the velocity as an effective thermal bath.  On the
other hand, in ABP, $\mathbf{k}_i$ is perpendicular to $\mathbf{n}_i$,
the orientation of the active force, and is a multiplicative noise
since its amplitude depends both on $\mathbf{v}_i$ and $\mathbf{F}_i$
through the unit vector $\mathbf{n}_i$. 
Also notice that (since
$\xi_i$ has unit variance and $\mathbf n$ is a unit vector) the
variances of the ABP and AOUP noises coincide, being $v_0^2/\tau$.  We
underline that, upon fixing $v_0$, the noise variance decreases in the
large persistence regime.

\subsection{Known results and positional order}\label{appendix:structuralproperties}

In the present Section, we illustrate the phase diagram for the values
of the control parameters $\phi$ and $\tau$ (at fixed $v_0$)
explored in this work.  Our results are obtained by means of
numerical solutions of Eqs.~\eqref{eq:x_dynamics_ABP} in a square
domain of size $L$ under periodic boundary conditions. 
In the considered interval of packing
fractions ($\phi \in [0.78,1.1]$), a suspension of passive Brownian
particles exhibits liquid, hexatic and solid phase depending on the
temperature. Our ABP phase diagram is consistent with the findings of Di
Gregorio et al.~\cite{digregorio2018full}.  As shown by these
authors, as $\tau$ is increased the liquid-hexatic and the
hexatic-solid boundaries shift to larger values of $\phi$ and the
hexatic region of the phase diagram is enlarged with respect to the
passive equilibrium picture.

\begin{figure*}[!t]
\centering
\includegraphics[width=0.9\linewidth,keepaspectratio]{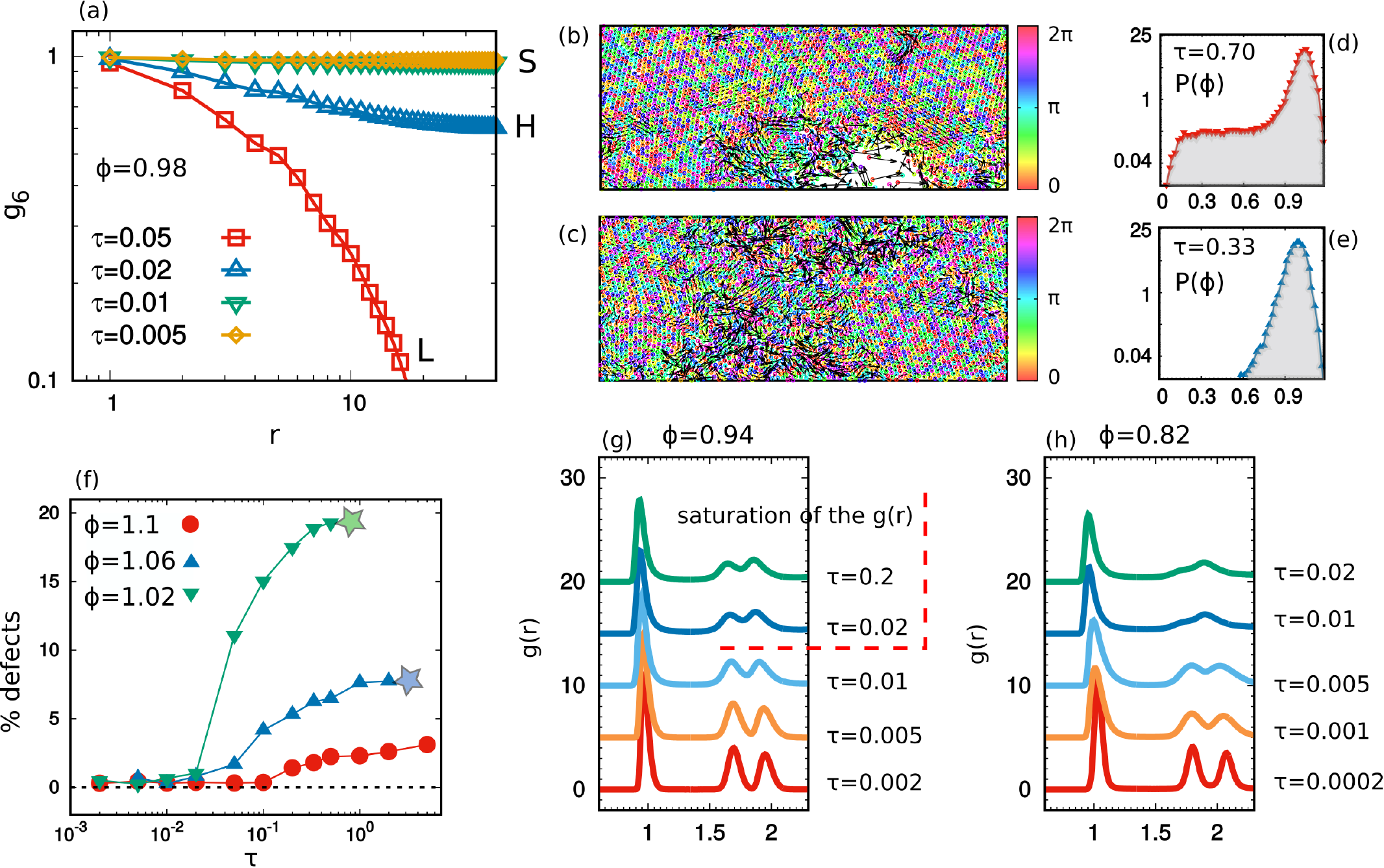}
\caption{ Structural properties. Panel (a):  $g_6(r)$ at $\phi=0.98$ for different values of $\tau$, as reported in the caption, exploring the different structural phases: the symbols S, H and L refer to the almost-solid, hexatic and liquid phases, respectively.
Panel (b), (c) are two snapshot
  configurations where colors denote the orientation of the particles,
  namely $\theta$, while panels (d) and (e) show the distribution of
  the packing fraction, $P(\hat{\phi})$, corresponding to the two configurations.
  In panel (f), we study the fraction of defects in the whole box vs
  $\tau$ for three different values of the densities as illustrated in
  the caption. The colored stars sign the value of $\tau$ at which the
  system becomes non-homogeneous.  Panels (g) and (h) show a set of
  $g(r)$ obtained at different $\tau$ for $\phi=0.94, 0.82$,
  respectively. Each curve is shifted along $y$ for presentation
  reasons.  Simulations are always realized with $v_0=50$ and the
  interaction discussed in the text.  }
\label{fig:phasediagramdensity}
\end{figure*}

When the persistence is smaller than the time-scale associated with
the potential, i.e. when $\tau \ll (\nabla\cdot \mathbf{F}(\bar{x})
/\gamma)^{-1}$, 
(being $\bar{x}$ the average distance between neighboring particles, which is fixed by the density in any homogeneous configurations)
we expect the same behavior as passive Brownian
particles~\cite{caprini2019activityinduced}: particles are
homogeneously distributed in the box and arranged in the solid,
hexatic or liquid phases, as shown in~\cite{digregorio2018full},
depending on the interplay between $\phi$ and $\tau$. In this regime, the active
force acts as a thermal bath with effective diffusivity $\sim v_0^2\tau$. 
Thus, the growth of $\tau$, at fixed $v_0$, can be mapped in the increase of the effective diffusivity in the corresponding Brownian system.
Depending on the interplay between $\phi$
and the effective temperature, the system explores liquid, hexatic or
solid phases. 
The structural properties are detected monitoring the
behavior of the orientational order parameter~\cite{cugliandolo2017phase, cugliandolo2018phases}, $\psi_6\left(\mathbf{x}_{i}\right)$, defined as $\psi_6\left(\mathbf{x}_{i}\right) = \sum_j
e^{6i\alpha_{ij}}/N_i$, where $\alpha_{ij}$ is the angle - with respect to
$x$ axis - of the segment joining the $i$-th and the $j$-th particle and the sum is restricted to the first neighbors of the particle $i$, namely $N_i$. 
In particular, we focus on the correlation function, $g_6(r=|\mathbf{x}_{i}-\mathbf{x}_{j}|)=\langle
\psi_6(\mathbf{x}_i)\psi_6^*(\mathbf{x}_j)\rangle/\langle \psi_6^2(\mathbf{x}_j)\rangle$, to distinguish between different phases.  
While $g_6(r)$ is roughly constant with
the distance in the solid phase, it decays as an inverse power-law in the hexatic phase.
Differently, in the liquid phase, $g_6(r)$ shows
an exponential decay. Examples of the different decays of $g_6(r)$ in
the three structural phases are shown in
Fig.~\ref{fig:phasediagramdensity}~(a).

In addition, we also measure the pair correlation function, $g(r)$.
At high density, the arrangement of particles is close to the hexagonal
lattice, so that the peaks of the $g(r)$ are placed at positions
$\bar{x}, \sqrt{3}\bar{x}, 2\bar{x}, \sqrt{7}\bar{x}, ... $ and so on,
where $\bar{x}$ is the typical distance between neighboring particles,
fixed by the density in any homogeneous configurations. It is worthy
to note that $\bar{x}$ is quite smaller than $\sigma$, meaning that
particles climb on the interacting potential due to the large
densities.  Each particle only interacts with its six neighbors due to
the truncated nature of $U(r)$.  In analogy with an equilibrium
system, one can roughly identify the liquid phase with the parameter
region where the second peak of $g(r)$ is not split.  We see that for
all values of the packing fraction, the peaks of the $g(r)$ decrease
as $\tau$ is increased, in agreement with the interpretation of the
self-propulsion in terms of effective diffusivity: fluidization
occurs.  The $g(r)$ is shown in panels (g) and (h) of
Fig.~\ref{fig:phasediagramdensity} for several values of $\tau$ and two
densities $\phi=0.94, 0.82$, respectively: at high density (g), the
system maintains a split second peak, while at moderate density (h),
the system shows a transition - increasing $\tau$ - towards a single
second peak where a liquid-like structure occurs~\cite{caprini2019activityinduced}.  Increasing
$\tau$ beyond some threshold value (that depends on $\phi$) denoted as
a red dotted line in panel (g), the curves representing $g(r)$
saturate, meaning that the internal structure is no more influenced by
the value of $\tau$.  On the other hand, the shape of $g(r)$ changes
towards less fluid configurations where the system shows
phase-separation or inhomogeneities. This is not a surprise since, in
those cases, particles in the clusters attain a more compact
configuration.

The boundary line between the homogeneous and inhomogeneous
(phase-separated) regimes is obtained by monitoring the distribution
of the local packing fraction, $P(\hat{\phi})$, shown in panels (d)
and (e) of Fig.~\ref{fig:phasediagramdensity} for two different
configurations at the same densities and different $\tau$. When the
system is spatially homogeneous $P(\hat{\phi})$ displays small tails
and a peak located at $\hat{\phi}=\phi$, while in the inhomogeneous
region it displays a long tail for small $\hat{\phi}<\phi$ and a shift
of the main peak for $\hat{\phi}>\phi$. 
The line of the transition from homogeneous (L, H, S) to phase-separated (P-H) phases is
tracked in Fig.~\ref{fig:velocitysnapshots_extreme}~(a), in correspondence of the first point showing such a shift.

Finally, the fraction of defects vs $\tau$ is measured for the denser configurations spanning solid and hexatic phases, as illustrated in
Fig.~\ref{fig:phasediagramdensity}~(f).
A defect is detected by counting
the number of neighbors of a particle inside a circular radius of size
$\sigma$: if the number of neighbors is different from six we mark
this point as a defect.   
The measure is
stopped when the system becomes inhomogeneous, in the proximity of colored stars. 
We observe that the solid-hexatic transition
takes place where the fraction of defects reaches $\sim 5\%$.

\section{Order in the velocities}\label{Sec:highdensities_numerics}

\begin{figure*}[!t]
\centering
\includegraphics[width=0.95\linewidth,keepaspectratio]{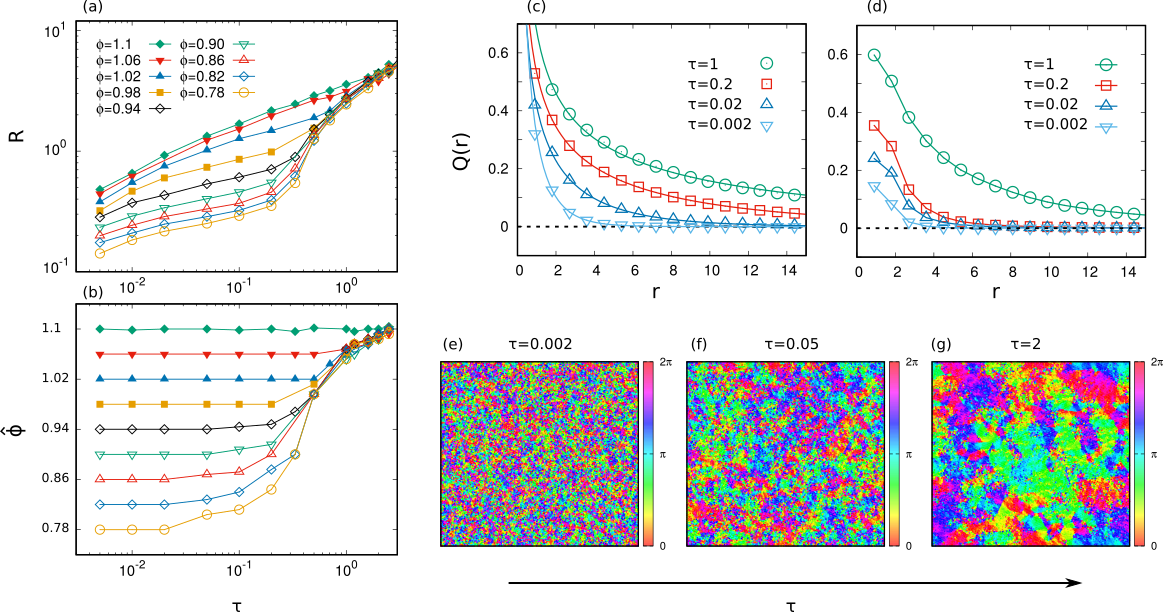}
\caption{ Size of the velocity aligned domains. In panels (a) and (b),
  we plot $R$ and $\phi$ as a function of $\tau$ for different values
  of ${\phi}$ as indicated in the caption.  Panels (c) and (d) show the
  spatial correlations of the velocity orientation, $Q(r)$, for
  ${\phi}=1.1, 0.94$, respectively. Curves with different $\tau$ are
  reported according to the caption. Panels (e), (f) and (g) show
  three different snapshot configurations for
  $\tau=0.002, 0.05, 2$, respectively, with $\phi=1.1$. All data in
  this figure are obtained at $v_0=50$. 
}
\label{fig:Qextreme}
\end{figure*}

Fig.~\ref{fig:velocitysnapshots_extreme}~(c) and~(d) are snapshots of
the system representing particles' positions and velocities for a large value of the persistence: 
they do not show MIPS,
since at ${\phi}=1.1$ the density remains homogeneous in the
considered range of $\tau$.  In the case of interacting systems, the
velocities of the particles, $\mathbf{v}_i$, represented by black
arrows in Fig.\ref{fig:velocitysnapshots_extreme}~(c)-(d), differ from
$\mathbf{f}^a_i$ and, in spite of the absence of any alignment
interactions, align and self-organize in large oriented domains. On
the contrary, the self-propulsion $\mathbf{f}^a_i$ remains randomly
oriented. 
The alignment of velocity orientations corresponds to the collective
movements of large domains of particles. Such domains rearrange
continuously in time and, sometimes, collapse into vortex structures, at
variance with the well-known traveling bands occurring in the
Vicsek-like models~\cite{gregoire2004onset, chate2008collective,mishra2010fluctuations, menzel2012collective}.

Hereafter, such a velocity order is studied quantitatively in terms of
spatial alignment velocity correlations and suitable order parameters,
useful to estimate the size of the domains. 
We find that there are different aspects in the velocity ordering phenomenology: i) order in the orientation of the velocity vectors ii) order in the full velocity vectors, accounting also for the occurrence of large regions with the same kinetic energy.

\subsection{Velocity orientation}

For Vicsek-like models, the global alignment of the particles, also
known as polarization, is commonly measured by the following order
parameters \cite{vicsek1995novel, vicsek2012collective, cavagna2014bird, chate2008collective}:
\begin{equation}
\label{eq:Vicsekorderparameter}
\varphi=\frac{1}{N}  \left| \sum_{i=1}^N e^{i\Theta_k} \right| \,,
\end{equation}
where $i$ is the imaginary unit and $\Theta_k$ is the velocity
orientation of the $k$-th particle.  This observable is almost zero
for particles without any alignment, typically at low numerical
densities and high noise, and returns one, for perfectly aligned
particles, e.g. for large values of $\phi$~\cite{chate2008collective}.
On the contrary in most systems of swimming active particles,
typically evolved by means of overdamped equations, velocity vectors
are ignored and the self-propulsion orientation is the only
information used to characterize polarization: 
for instance, in the case of spherical (apolar) ABP particles, the
parameter $\varphi$ in Eq.~\eqref{eq:Vicsekorderparameter} - with $\Theta_k$
replaced by $\theta_k$ of Eq. \eqref{eq:theta_dynamics} - is close to zero. 
In this model, since self-propulsions do not coincide with velocities, it is more suitable to consider 
the orientation of the velocity vector,
$\dot{\mathbf x}_k$, in Eq.~\eqref{eq:Vicsekorderparameter},  
i.e. replacing $\Theta_k$ with the angle formed by the
velocity of the particle with respect to the $x$ axis. 
However, due to the presence of
several domains with different orientations, there is no global
velocity-polarization.  In principle, for very large persistence we
could observe a large oriented domain spanning the whole box, but such
a finite size effect occurs only when $v_0 \tau \gg L$, i.e. when the
persistence length exceeds the size of the box~\cite{menzel2013traveling,menzel2014active}. 
We do not consider such a case in this manuscript.

A more appropriate indicator, which - even in the absence of a global
polarization - gives information about the local alignment of the velocities
and its dependence on physical parameters, is the spatial
correlation function of the velocity-orientation, $Q_i (r)$~\cite{caprini2019spontaneous}. 
This observable measures the velocity alignment between the particle $i$ and its
neighboring particles located in the circular crown of thickness
$\bar{r}= \sigma$  and mean radius $r=k\bar{r}$, being $k$ an integer positive
number, and reads: 
\begin{equation}
Q_i(r)=1- 2\sum_j \frac{d_{ij}}{\mathcal{N}_k \pi} \,
\end{equation}
where the sum runs over the particles within the circular crown
defined by the value of $k$ and $\mathcal{N}_k$ corresponds to the
number of particles contained in it.  The term $d_{ij}$ is the angular
distance between the two angles of the velocities of particles $i$ and
$j$, namely $\beta_i$ and $\beta_j$, calculated as 
$d_{ij}=\text{min}\left[ |\beta_i - \beta_j|, 2\pi - |\beta_i - \beta_j| \right]$.
We average over all particles by
defining $Q(r) = \sum_i Q_i(r)/N$ which has the property of being $1$
and $-1$ for perfectly aligned and anti-aligned particles,
respectively, and $0$ in the absence of any form of alignment.  In panels
(c-d) of Fig.\ref{fig:Qextreme}, we report $Q(r)$ for different values
of the persistence time, $\tau$, and for two different densities. As
expected $Q(r)$ is a decreasing function of $r$.  For the smallest
values of $\tau$, the alignment is appreciable only in the first
shells, a finding consistent with the scenario where the self-propulsion only acts as an effective thermal diffusion.
Instead, as $\tau$ increases, $Q(r)$ takes on larger values in the
first shell and decays slower and slower towards zero, with a typical
decay length which roughly represents the average size of one domain: larger
values of $\tau$ and density (almost always) produce both the
increasing of $Q(r)$ in the first shell ($k=1$) and the slower decay
of the whole function.  This observation is, also, qualitatively
confirmed by three snapshot configurations obtained for increasing
values of $\tau$ from panel (e) to panel (f) and (g) of
Fig.~\ref{fig:Qextreme}. There, the colors encode the velocity
orientations, showing the growth of the average size of the individual
velocity domains with $\tau$.

The velocity ordering is well captured by the order parameter $R$,
obtained by integrating over the whole box the correlation $Q(r)$:
\begin{equation}
R=\int Q(r) dr \, .
\end{equation}
Such a parameter is a measure of the domain size and is studied  
in Fig.~\ref{fig:Qextreme}~(a), as $\tau$ varies for
several values of $\phi$, evaluating both the homogeneous and the
non-homogeneous configurations. We recall that the non-homogeneous
regimes correspond to the emergence of empty regions or
phase-separation, signaled by the presence of a non-single-peak in the
packing fraction distribution (see Fig.~\ref{fig:phasediagramdensity}~(d-e)). 
For the purpose of evaluating the impact of density
inhomogeneity upon $R$, we show the main peak position, $\hat{\phi}$
in Fig.~\ref{fig:Qextreme}~(b), for different values of $\tau$ and
average packing fraction $\phi$. We remark that $\hat{\phi} \approx
\phi$ up to some critical value of $\tau$, then it increases. Such a
critical value grows as the average packing $\phi$ is increased. At the larger packing fraction studied $\phi=1.1$ the system
remains homogeneous for all the explored values of $\tau$.  It is
worthy to note that the values of $\hat{\phi}$ (the densities in the
denser portion of the system) become independent from $\phi$ for
$\tau$ sufficiently large.

\begin{figure*}[!t]
\centering
\includegraphics[width=1\linewidth,keepaspectratio]{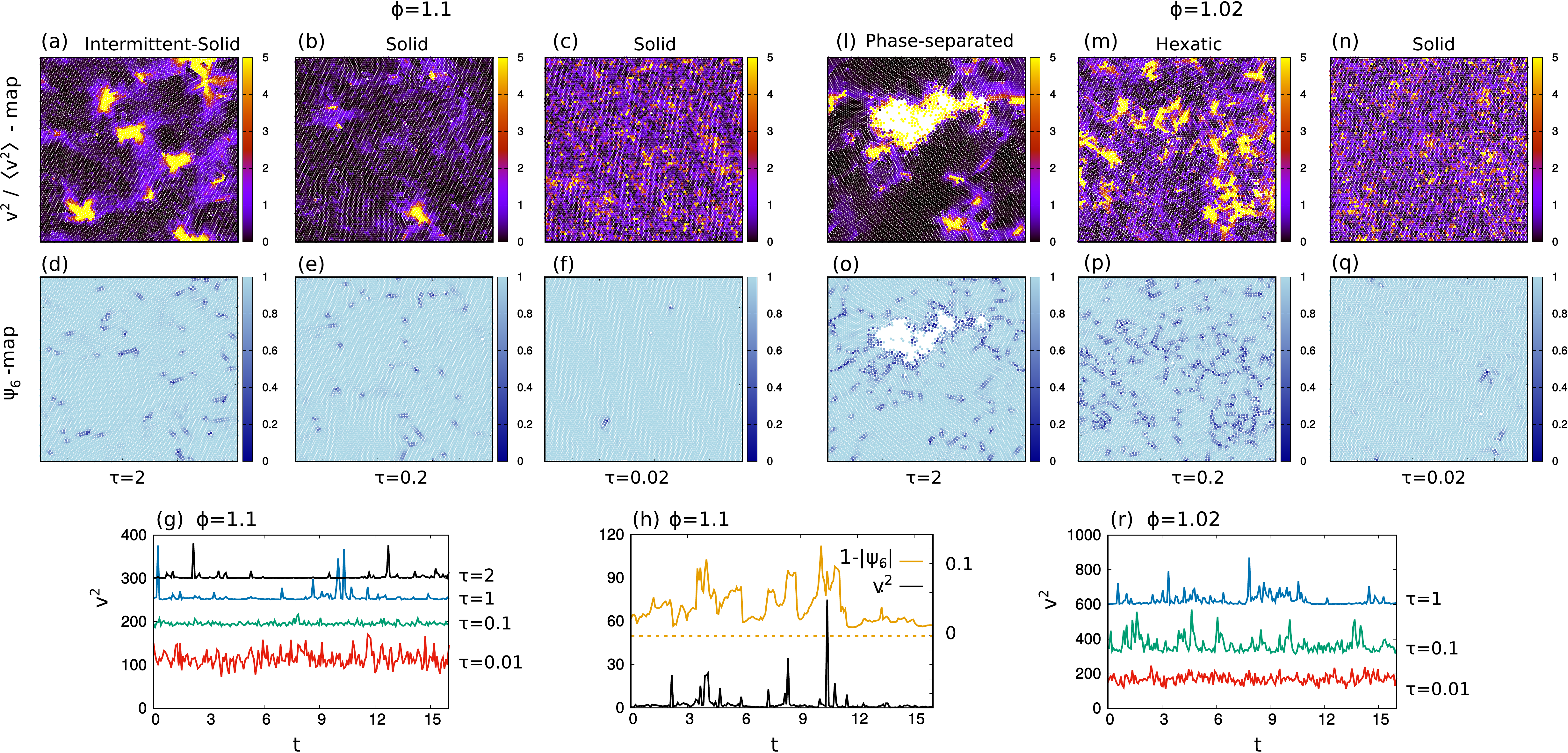}
\caption{ Fluctuations of kinetic energy. Panels (a)-(f) and (l)-(q)
  are snapshots in the plane $xy$ realized at $\phi=1.1, 1.02$,
  respectively, for different values of $\tau$ as reported below the
  panels.  Colors of panels (a), (b), (c) and (l), (m) and (n) encode
  the value of the square velocity of each particle, $v^2/\langle v^2\rangle$, while colors of
  panels (d), (e), (f) and (o), (p), (q) represent the value of the
  crystalline orientational order parameter $|\Psi_6|$.  Panels (g)
  and (r) report different time-series of $v^2$ obtained for
  $\tau=10^{-2},10^{-1}, 1, 2$ as indicated in the caption, with
  $\phi=1.1$ (panel (g)) and $\phi=1.02$ (panel (r)).  Panel (h)
  compares a single trajectory of $v^2$ and $1-|\Psi_6|$, obtained
  with $\tau=2$ and $\phi=1.1$.  All the simulations are realized with
  $v_0=50$.  }
\label{fig:Fig}
\end{figure*}

As shown in panel (a), $R$ increases with $\tau$.  For $\phi=1.1$,
i.e. when the system is spatially homogeneous for all the values of
$\tau$, the growth is steady.  This proves that the growth of $R$ occurs even in the
absence of any local density inhomogeneity since it is not associated with some local change of density.  
Instead, for smaller values of
$\phi$, a first slow monotonic increase is followed by a sharp one
occurring at a value of $\tau$ for which the homogeneous liquid-like
or hexatic phases break down in favor of an inhomogeneous phase.  The
comparison between panels (a) and (b) also suggests that $R$ has a
strong dependence on $\phi$.

The occurrence of such a velocity order is an evidence of the
non-equilibrium nature of the dense active phases.  Even if the
structural (positional) information suggests an analogy with the liquid,
hexatic or solid phases of a passive - equilibrium - system, there is
not an equilibrium counterpart of the velocity ordering phenomenon.

In general, it is believed that the occurrence of local velocity alignment
is a consequence of the breaking of some microscopic isotropy (as occurs in the case of rod or elongated particles~\cite{peruani2006nonequilibrium, ginelli2010large, abkenar2013collective, peruani2017hydrodynamic}) or the introduction of explicit alignment interactions.
The present ABP model subject to random independent active driving leads
to velocities' alignment  - in the high-density regimes - even for spherical particles.
This phenomenology simply arises from the interplay between self-propulsion and steric inter-particle repulsion.

\subsection{Kinetic energy and intermittency}

Besides the local order of velocity orientations, spatial correlations
also manifest in the speed, $v=|\mathbf{v}|$, and, thus, in the
kinetic energy of the particles, $\propto v^2$.  Panels (a), (b) and
(c) of Fig.~\ref{fig:Fig} show the map of $v^2/\langle v^2\rangle$ for three snapshot
configurations obtained varying $\tau$, for $\phi=1.1$, i.e. at a
density value such that the active system attains a solid-like state
for every $\tau$.  For small $\tau$, kinetic energies display uncorrelated spatial fluctuations (with Gaussian statistics, not shown here). 
As $\tau$ grows, 
structures characterized by similar (or correlated) values of $v^2$
appear, with alternation of fast and slow regions (each identified by
a given color).

We also highlight an interesting connection between the kinetic energy spatial
distribution and the structural properties of the system. In panels
(d), (e) and (f) of Fig.~\ref{fig:Fig}, we plot the observable
$|\psi_6(\mathbf{x}_i)|$ - which is the field pertaining to the crystalline
orientational order - relative to the three configurations of panels
(a), (b) and (c).  The comparison between the maps of $|\psi_6(\mathbf{x}_i)|$ and
$v^2$ reveals that the regions with large kinetic energies develop
close to the defects of the crystalline structure.  A similar scenario
occurs for a smaller value of $\phi$, namely $\phi=1.02$.  In this
case, the $v^2$-map is shown in panels (l), (m) and (n), while panels
(o), (p) and (q) report the $|\psi_6|$-map.  For this choice of
$\phi$, the three values of $\tau$ distinguish between different
aggregation phases: phase-separated, hexatic and solid (from the left
to the right).  The solid phase for the smaller value of $\tau$ is
qualitatively indistinguishable from the denser case (compare the
panels (c) and (n)).  Instead, for the intermediate value of $\tau$
(panel m)) the occurrence of the hexatic phase  is responsible for a
larger number of defects and, thus, a larger number of
mobile particles, as clearly shown from the comparison between panels (m)
and (p).  Finally, in the phase-separated configuration, the fastest
regions are mostly concentrated near the boundary of the empty region
(panel (l)).

To have another perspective, it is instructive to consider the
time behavior of the kinetic energy, $v^2$, calculated averaging
over a box of size $l$ such that $\bar{r}\ll L$.  This observable, as a function of time, is reported
for different values of $\tau$ in Fig.~\ref{fig:Fig}~(g) and (r) for
$\phi=1.1, 1.02$, respectively.  In these two cases, the scenario is
similar: for the smaller values of $\tau$ the kinetic energy displays
symmetric and rapidly uncorrelated fluctuations around the mean
value, $\langle v^2 \rangle$. This is coherent with an effective
equilibrium picture which is expected when $\tau \to 0$.  For large
values of $\tau$, sparse anomalous peaks manifest, corresponding to
rare fluctuations, which move away from their average by several
standard deviations.  Such peaks become higher and more isolated when
$\tau$ increases. The observed behavior resembles the temporal
intermittency observed in turbulence~\cite{bohr2005}.  A
more detailed analysis of such an issue will be presented in
a future work, while, in this manuscript, we only consider the essential features of this
phenomenon, in particular the role played by defects in the solid
phase: particles near the defects attain - in fact - large kinetic
energy. To corroborate our observation, we compare the fluctuations of the orientational order parameter, $\Psi_6=\sum_{i=1}^{N_l} \psi_{6,j}/N_l$,
and those of kinetic energy, $v^2$, both averaged over a box of size $l$ with $N_l$ particles. 
In particular, Fig.~\ref{fig:Fig}~(h) shows the time-trajectories of $1-\Psi_6$ and $v^2$ revealing a fair correlation
between the occurrence of spikes for both these observables.

With the aim of introducing also the information about intermittency
in the general picture, we have drawn a dashed black line in the phase
diagram of Fig.~\ref{fig:velocitysnapshots_extreme}: the line
identifies the intermittency region and is tracked at the first values
of $\tau$ and $\phi$ for which the peaks of the $v^2$ trajectory
overcomes $3$ times the standard deviation from its average
value. 

\subsection{Vectorial velocity field}

\begin{figure*}[!t]
\centering
\includegraphics[width=1\linewidth,keepaspectratio]{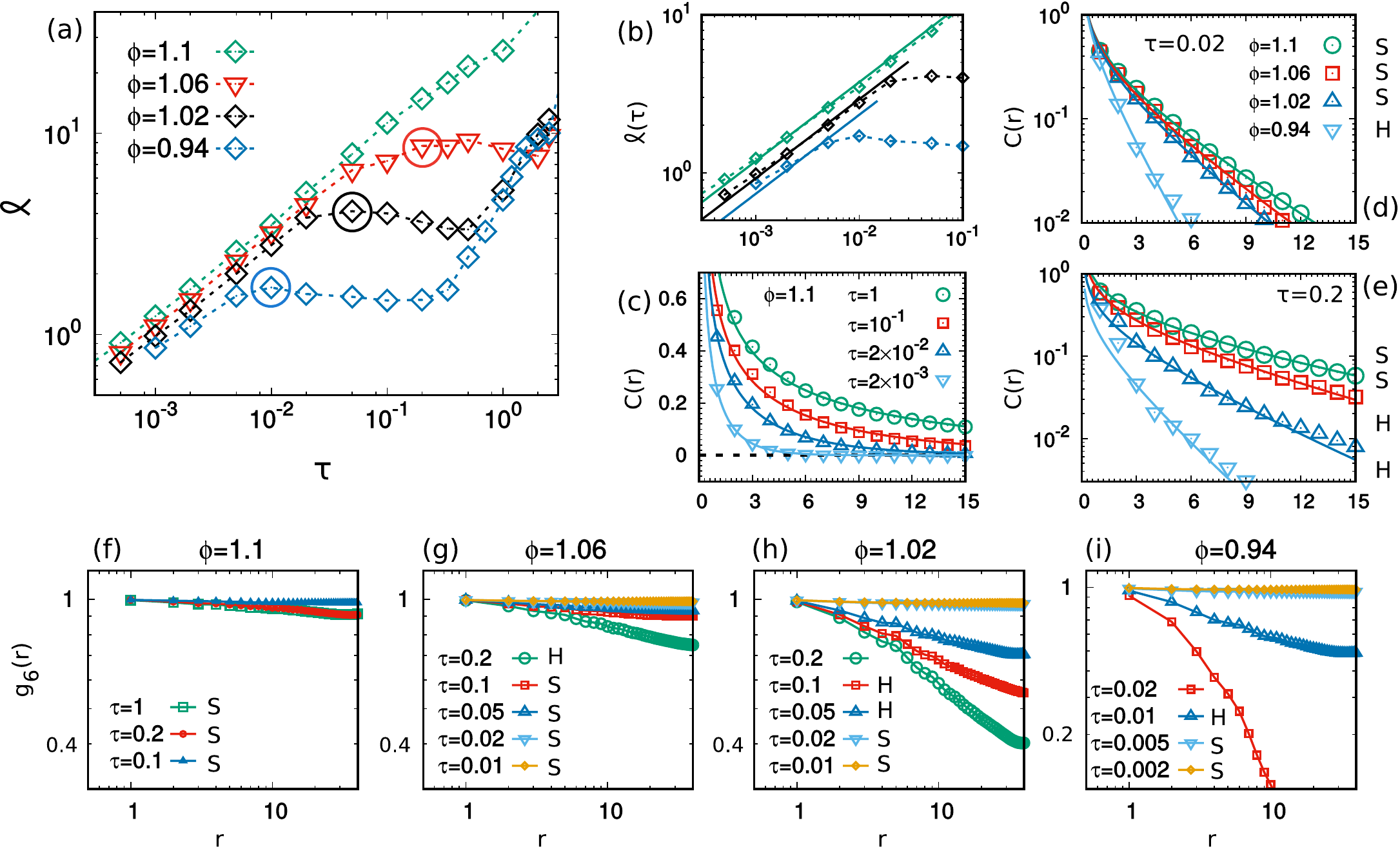}
\caption{ Spatial correlations of velocity vectors: theory and
  numerics. Panel (a) and (b) show the correlation lengths, $\ell$, as
  a function of $\tau$ for different values of $\phi$ as shown in the
  legend. The colored circle around a dot of each curve in panel (a)
  is the point at which the solid-hexatic transition is obtained.  In
  particular, panel (b) zooms into small values of $\tau$ and shows
  the prediction~\eqref{eq:lambdaprediction} as a solid line, fitting
  the function $a\,\tau^{1/2}$, being $a$ a fitting parameter.  Panels
  (c), (d) and (e) plot the correlation function $C(r)$ for several
  values of $\phi$ and $\tau$: for $\phi=1.1$ (i.e. in the solid- state) for several values of $\tau$ (panel (c)), and for two values of
  $\tau$ as a function of $\phi$ (panels (d) and (e) which share the same caption). 
  In panels (c), (d), (e), points are obtained from numerical simulations while solid lines from the theoretical prediction, Eq.~\eqref{eq:predictionofcorrelation}.
The letters S and H near the curves mean ``solid'' and ``hexatic'' phase,
  respectively.  Panels (f), (g), (h) and (i) report $g_6(r)$ for
  different values of $\phi$ and $\tau$.  The simulations are realized
  with $v_0=50$.
 }
\label{fig:corr}
\end{figure*}

In the previous Sections, we have seen that the spatial correlation of
velocity orientations grows with $\tau$, even in the presence of
crystalline defects or large voids (such as those in the
phase-separated regimes). Speed (velocity modulus) is more sensitive
to the presence of defects and creates patterns with sparse strong
fluctuations. A natural question arises: what happens to the spatial
correlation of the full velocity vectors (which incorporate both
orientation and modulus)?

A quantitative measure of the ordering of the velocities can be
obtained by measuring the spatial correlation function 
\begin{equation}
C(r)=\frac{\langle \mathbf{v}(r)\cdot \mathbf{v}(0)\rangle}{\langle v^2 \rangle} \,,
\end{equation}
in the continuous limit,
$\mathbf{v}_i \to \mathbf{v}(\mathbf{r})$.
$\langle v^2\rangle$ is the variance of the velocity distribution calculated over the whole box.
Our analysis limited to the case of homogeneous density
under some assumptions is able to predict the form of $C(r)$ (as illustrated in
Appendix~\ref{app:appendixderivation}).
  The equation of
motion~\eqref{eq:xv_dynamics_MIPS}  is approximated by the AOUP dynamics, replacing
the multiplicative noise by a two-dimensional additive noise. By this
method, it is possible to predict the spatial velocity correlation
function. At variance with a similar calculation reported
in~\cite{caprini2019spontaneous}, here we assume that particles
are free to oscillate around their equilibrium positions, i.e. they
form a hexagonal crystal structure with oscillating sites.
Under these simple hypotheses, an expression for
$C(r)$ can be derived using the equation of evolution of the velocities. 
In particular, $C(r)$ displays an exponential-like behavior:
\begin{equation}
\label{eq:predictionofcorrelation}
C(r) \propto  \frac{ \bar{x}^2}{\ell ^2}
\Bigl(\frac{\ell}{8\pi r }\Bigr)^{1/2}   e^{-  r/\ell} \,,
\end{equation}
where $\ell$ is the correlation length:
\begin{equation}
\label{eq:lambdaprediction}
\ell = \bar{x}\sqrt{\frac{\tau}{\gamma}}\left[ \frac{3}{4} \left(U''(\bar{r}) + \frac{U'(\bar{r})}{\bar{r}} \right)   \right]^{1/2}\,.
\end{equation}
Thus, a strong potential and/or a large value of $\tau$ increases the
value of $\ell$.  Also increasing the average packing
fraction (i.e. decreasing the lattice constant $\bar{r}$) leads to a
growth of $\ell$ through the $U(\bar{r})$ dependence
on this quantity.  Expression~\eqref{eq:lambdaprediction} coincides
with the result obtained in~\cite{caprini2019spontaneous}, even if
here it is derived under the less restrictive hypothesis of a
vibrating (not rigid) lattice. 
Moreover, at variance with~\cite{caprini2019spontaneous}, here we are also considering very high
average densities with homogeneous (not phase-separated)
configurations, allowing us to directly check the scaling of $C(r)$
with $\tau$.  
On the contrary, when phase-separation occurs, the
packing fraction $\hat{\phi}$ of the dense regions (and therefore the
effective value of $\bar{r}$) grows with $\tau$, even at fixed
average density $\phi$. 

To check the predictions~\eqref{eq:predictionofcorrelation} 
and~\eqref{eq:lambdaprediction}, we study the spatial velocity
correlation, $C(r)$, for several values of $\tau$ and $\phi$.  $C(r)$
for the denser case ($\phi=1.1$), which corresponds to the solid phase
for the whole range of $\tau$ numerically explored, is reported in
Fig.~\ref{fig:corr}~(c) and reveals a good agreement with
theory.  The correlation length, $\ell$, is reported in
Fig.~\ref{fig:corr}~(a) (green diamonds) and shows a monotonic
growth in fairly good agreement with Eq.~\eqref{eq:lambdaprediction}:
under these high-density conditions, the defects are not so
statistically relevant and do not interfere with the velocity order.
A zoom of Fig.~\ref{fig:corr}, shown in panel (b) and accompanied by a fit of the numerical data, 
displays a good agreement with  formula~\eqref{eq:lambdaprediction}.  
A similar analysis reveals discrepancies between theory
and simulations when applied to lower values of $\phi$: 
the growth of $\ell$ ceases at some value of $\tau$
which depends on $\phi$. We mark with a colored circle the first value
of $\tau$ where $\ell$ has reached a plateau, and call it
$\tau^*(\phi)$.  Interestingly, this is close (up to numerical errors)
to the value of $\tau$ where the solid-hexatic transition takes place
(i.e. the fraction of crystalline defects roughly overcomes a given threshold),
Here, for completeness, we report the correlation function of
$\Psi_6$, namely $g_6 = \langle \Psi_6(0) \Psi_6^*(r) \rangle$, in
Fig.~\ref{fig:corr}~(f)-(i).
This function shows the
well-known transition from the solid to the hexatic phases, roughly at
$\tau^*$, where $g_6$ goes from a nearly constant behavior to a power-law decay~\cite{cugliandolo2018phases}.
For comparison, we also show $C(r)$ as a function of $\phi$ for
two different values of $\tau$, in panels (d) and (e), where each phase is signed by S, H (solid, hexatic) in the caption.
In the hexatic
phase, $C(r)$ maintains the exponential-like shape predicted in Eq.~\eqref{eq:predictionofcorrelation} but decays abruptly much faster in the proximity of the solid-hexatic transition.
As a consequence, the agreement between Eq.~\eqref{eq:lambdaprediction} and
data, shown in Fig.~\ref{fig:corr}~(b), holds up to $\tau^*$, while for
$\tau > \tau^*$ the presence of defects invalidates the
prediction~\eqref{eq:lambdaprediction} since $\ell$ remains nearly constant or decay very slowly.
It is also remarkable that
a further increase of $\tau$ produces a steep growth of $\ell$.
This is likely caused by phase-separation and the increase of
local density in the clustered regions.
Even in this case, we still expect that Eq.~\eqref{eq:lambdaprediction} holds even if
the function $\bar{x}(\tau)$ is unknown.

In conclusion, we have solid
arguments to state that a large number of defects, occurring in the
hexatic or liquid phase, is responsible for the saturation of $\ell$.
Indeed, in the proximity of a defect, regions with large kinetic energies are
present, as seen in the previous section, and, as a consequence, the
velocities are less correlated. 
Interestingly, the size $R$ of the orientational domains is not so affected by the lack of orientational order, always revealing a monotonic growth with $\tau$ independently of the structural phase.

\section{Discussion and perspectives}\label{sec:conclusion}

We have studied systems of self-propelled particles at high packing
fractions displaying structural properties which resemble the
equilibrium liquid, hexatic and solid phases, exploring both the small
and the large persistence regime.

While at small values of the persistence time, $\tau$, many observables behave as
in equilibrium, the phenomenology in the high
persistence regime is much richer and displays unexpected
phenomena. We conclude that an improved active phase diagram
benefits from the introduction of new order parameters, related to
velocity correlations which have not a Brownian counterpart.

In particular, velocities exhibit intriguing patterns, forming aligned
domains or vortex-like arrangements.  We propose a suitable order
parameter which quantifies the size of these domains, deduced from the
spatial correlations of the velocity orientations.  Such a parameter
grows with both packing fraction, $\phi$, and $\tau$ in the homogeneous phases, but it
becomes independent from $\phi$ in the phase-separated regimes.  We
also observe the occurrence of large regions whose kinetic energy is
highly correlated: these regions become larger when $\tau$ and $\phi$
grow. Besides, high energetic regions form in the proximity of
orientational defects of the solid phase, much more visible in the
hexatic phase where defects are diffuse.  This is accompanied by a
pronounced time intermittency phenomenon, apparently well-correlated
to the fluctuations of the orientational order parameter. Correlations
of the full velocity vector are also useful to get insights about
these dynamical features, in particular, they can be successfully
compared to a mesoscopic theory developed under the assumption of
homogeneous density. Deviations from the theory appear together with
the emergence of a large fraction of defects and the breakdown of homogeneity.

Our observations call for experimental verifications.  Promising real
platforms to confirm such interesting phenomenologies are
Janus particles~\cite{takatori2016acoustic, klongvessa2019relaxation}
or vibrated polar granular disks~\cite{briand2016crystallization}.

In the present work, we have been interested in understanding the dense phase in monodisperse
active systems. Binary mixtures are often employed for gaining insight into active glass phases~\cite{berthier2011theoretical,berthier2019glassy,janssen2019active,szamel2015glassy, berthier2017active,berthier2019glassy,ni2013pushing,mandal2016active,nandi2018random}. In particular,
it has been shown that the spatial velocity correlation function is an input ingredient for developing a self-consistent Mode-Coupling Theory of Active Matter~\cite{szamel2015glassy}.
Our findings prove 
that, as a general result, the statistical properties of the velocity field have to be taken into account 
for providing a complete description of active systems. 
As a future direction, it would be interesting to understand how the velocity alignment patterns we observed 
modify glassy transition.

The phenomenology of domains with aligned velocities could resemble the scenario of
traveling crystals,  
observed in Refs.~\cite{menzel2013traveling,
  menzel2014active} in numerical simulations.
  Recently, traveling crystals have been observed in experiments
  using suspensions of micro-disks subjected to vertical vibrations~\cite{briand2018spontaneously}.  
In such studies, the whole hexagonal
pattern moves coherently in space, even though each
self-propulsion vector points randomly.  
Our phenomenology is quite different since far particles (which belong to different domains) move
independently and, thus, the whole crystal gets stuck. 
Instead, the movement of some clusters gives rise to the
formation of defects.  We find that the whole crystal travels coherently as in 
Ref.~\cite{menzel2013traveling} only if the size of the box is smaller than
the persistence length of the active motion.

\subsection*{Acknowledgements}
LC thanks M. Cencini and A. Cavagna for fruitful discussions.
AP and MP acknowledge funding from Regione Lazio, Grant Prot. n. 
85-2017-15257 ("Progetti di Gruppi di Ricerca - Legge 13/2008 - art. 4").
LC, UMBM, and AP acknowledge support from the MIUR PRIN 2017 project 201798CZLJ.

\appendix
\section{Velocity dynamics in the case $T>0$}\label{appendix:changeofvariablesT}

The change of variables from $(\mathbf{x}_i, \mathbf{f}^a_i)$ to $(\mathbf{x}_i, \mathbf{v}_i)$, i.e. from Eqs.~\eqref{eq:x_dynamics_ABP} and~\eqref{eq:theta_dynamics} to Eqs.~\eqref{eq:xv_dynamics_MIPS_varx} and~\eqref{eq:xv_dynamics_MIPS}, has been derived in Ref.~\cite{caprini2019spontaneous} in the absence of thermal noise (i.e. for $T=0$). The change of variables can be easily generalized to the case $T>0$, as shown in this Appendix. 
Such a generalization follows the strategy of Ref.~\cite{caprini2018active}, where the result is obtained in the case of the AOUP model for non-interacting particles. 
Here, the same trick can be adapted to the ABP self-propulsion.
To overcome the difficulty regarding the time-derivation of the thermal noise, we define the variable $\mathbf{v}_i=\dot{\mathbf{x}}_i-\sqrt{2 T/\gamma}\,\boldsymbol{w}_i$. Taking the derivative with respect to the time, the dynamics reads:
\begin{flalign}
\dot{\mathbf{x}}_i &= \mathbf{v}_i + \sqrt{2T/\gamma}\, \boldsymbol{w}_i \\
\label{eq:xv_dynamics_MIPSb}
\tau\gamma\dot{\mathbf{v}}_i &= - \gamma\sum_{j=1}^N{\Gamma}_{ij}({\mathbf r}_{ij}) \mathbf{v}_j + \mathbf{F}_i +  \tau\gamma\mathbf{k}_i\\
&\qquad\qquad\qquad - \tau\nabla_i \cdot \mathbf{F}_i \sqrt{2 T/\gamma}\boldsymbol{w}_i\,. \nonumber 
\end{flalign}  
The thermal noise comes into play with two additional noise terms. The first is an additive noise on the dynamics of $\dot{\mathbf{x}}_i$, while the second is multiplicative and acts on $\dot{\mathbf{v}}_i$.
Its space prefactor balances the space-dependent Stokes force in the equilibrium limit $\tau \to 0$.

\section{Positional, orientational and velocity correlations for a perfect active lattice}\label{app:appendixderivation}

\newcommand{\betadef}{\frac{1}{\tau}}
\newcommand{\alphadef}{\frac{\omega_q^2}{\gamma}}
\newcommand{\br}{{\bf r}}
\newcommand{\bu}{{\bf u}}
\newcommand{\bR}{{\bf x}}
\newcommand{\bRz}{{\bf x}^0}
\newcommand{\bk}{{ \bf k}}
\newcommand{\bx}{{ \bf x}}
\newcommand{\vv}{{\bf v}}
\newcommand{\nb}{{\bf n}}
\newcommand{\mb}{{\bf m}}
\newcommand{\bq}{{\bf q}}
\newcommand{\rb}{{\bar r}}

\newcommand{\eeta}{\boldsymbol{\eta}}
\newcommand{\xxi}{\boldsymbol{\xi}}

In order to obtain the correlation functions of the two-dimensional system we shall make two simplifying assumptions in the dynamics, Eq.~\eqref{eq:x_dynamics_ABP}:
\begin{itemize}
\item[i)] each component of the active force, $\mathbf{f}^a_i$, is replaced by independent Ornstein-Uhlenbeck processes, namely $\boldsymbol{\eta}_i$, with equivalent intensity, $v_0=\sqrt{D/\tau}$ and persistence time, $1/D_r=\tau$;
\item[ii)] each particle oscillates around a node of a hexagonal lattice so that the 
total inter-particle potential can be approximated as the sum of quadratic terms.
\end{itemize}
The approximate dynamics reads:
\begin{eqnarray}
&&
\dot \eeta_i(t) =-\frac{1}{\tau}\eeta_i(t)+ \frac{\sqrt{2 D} }{\tau} \xxi_i(t)\\&&
\dot \bR_i(t) =  \eeta_i(t)  
-  \sum_j^{n.n}\frac{\nabla_i U(|\bR_j-\bR_i|)}{\gamma} \,,
 \label{dynamicequation0}
\end{eqnarray}
where $\nabla_i U$  stands for the gradient of the potential $U$ with respect to $\bR_i$ and the sum involves the nearest neighbors
of the lattice node $i$.

Introducing the displacement $\bu_i$ of the particle $i$ with respect to its equilibrium position, $\bRz_i$, namely
\begin{equation}
\bu_i=\bR_i-\bRz_i \,,
\end{equation}
we get
\begin{eqnarray}
&&
\dot \eeta_i(t) =-\frac{1}{\tau}\eeta_i(t)+ \frac{\sqrt{2 D} }{\tau} \xxi_i(t)\\&&
\dot \bu_i(t) =  \eeta_i(t)  
+\frac{K}{\gamma} \sum_\mb^{n.n} (\bu_j-\bu_i) \,,
 \label{dynamicequation2}
\end{eqnarray}
where $K$ is the strength of the potential in the harmonic approximation, i.e. $U\approx\frac{K}{2} (\bu_j-\bu_i)^2$, which reads
$$
2 K = \left(U''(a) + \frac{U'(a)}{a} \right) \,,
$$
being $a=\bar{x}$ the lattice constant.
In order to solve the problem, we switch to normal coordinates, employing 
the Fourier space representation:
\begin{eqnarray}
&&
 \hat \bu_{\bq}=\frac{1}{ N}\sum_i   \bu_i\,  e^{-i \bq\cdot  \bRz_i }
\\&&
\hat \eeta_{\bq}=\frac{1}{ N}\sum_i   \eeta_i\,  e^{-i \bq\cdot  \bRz_i } \,,
\label{fourierrepresentation}
\end{eqnarray}
and obtain
\begin{eqnarray}
&&
\frac{d}{dt}\hat{\eeta}_\bq(t) =-\frac{1}{\tau} \hat \eeta_\bq+\frac{\sqrt{ 2 D}}{\tau} \hat \xxi_\bq \\&&
\frac{d}{dt}\hat \bu_\bq(t) =-\frac{\omega^2_\bq}{\gamma} \hat \bu_\bq(t) + \hat \eeta_\bq \,,
\label{dynamicequation4}
\end{eqnarray}
where
\begin{flalign}
\label{eq:app_omegaq}
\omega_\bq^2&=-2 K \Bigl[\cos(q_x a) +2\cos\Bigl(\frac{1} {2} q_x a\Bigr)\cos \Bigl(\frac{\sqrt 3} {2} q_y a\Bigr)-3\Bigr] \nonumber\\
& \approx \frac{3}{2} K a^2 q^2 + O(q^4)\,,
\end{flalign}
where $\mathbf{q}=(q_x, q_y)$ are vectors of the reciprocal Bravais lattice.
Thus, we can easily calculate the steady-state equal time correlations:
 \begin{eqnarray}
 &&
 \langle  \hat{\bu}_\bq(t) \cdot \hat{\bu}_{-\bq}(t)  \rangle= 
 \frac{2D\gamma}{\omega_\bq^2 \left(1+\frac{\tau}{\gamma}\omega_\bq^2 \right)}\\&&
\langle \hat{\vv}_\bq(t) \cdot \hat{\vv}_{-\bq}(t)  \rangle=\frac{2D}{\tau}
  \frac{1}{1+\frac{\tau}{\gamma}\omega_\bq^2}\\&&
\langle  \hat{\bu}_\bq(t)  \cdot \hat{\vv}_{-\bq}(t)  \rangle=0
\end{eqnarray}


\subsection{Velocity correlation function}
We, now, consider the real-space velocity correlation function:
\begin{equation}
\langle \vv_{\bf x} \cdot \vv_{\bf x'}\rangle
=\frac{1}{N^2}  \frac{2D}{\tau} \sum_{\bq} e^{i\bq (\bx-\bx')}\frac{1}{ (1+\frac{\tau}{\gamma}\omega_\bq^2)}  \,.
\end{equation}
By replacing the lattice sum by a double dimensional integral and defining $r=| {\bf x} - {\bf x'}|$, we have
\begin{equation}
\langle \vv_{\bf x} \cdot \vv_{\bf x'}\rangle \approx
 \frac{1}{2\pi}  \frac{2 D}{\tau} \frac{ a^2 }{\ell^2} 
  K_0(r/\ell)
 \end{equation}
 where $K_0(r/\ell)$ is the zero-order modified Bessel function of the second kind which has the following asymptotic behavior when $r/\ell\gg1$:
$$
K_0(r/\ell) \approx \Bigl(\frac{\pi \ell}{2  r}\Bigr)^{1/2}  e^{-  r/\ell} \,,
$$
we find
\begin{equation}
\langle \vv_{\bf x} \cdot \vv_{\bf x'}\rangle \approx   2 v_0^2  \frac{ a^2}{\ell ^2}
\Bigl(\frac{\ell}{8\pi r }\Bigr)^{1/2}   e^{-  r/\ell} \,,
\end{equation}
 where
$$
\ell^2= \frac{3\tau}{2\gamma } a^2 K  =\frac{3\tau }{4 \gamma} a^2 \Bigl( U''(a)+ \frac{U'(a)}{a}\Bigr) \,,
$$
which defines the correlation length $\ell$ in the harmonic hexagonal lattice in agreement with the result \eqref{eq:lambdaprediction}.

\subsection{Bond angle order and $\psi_6$-field in the harmonic crystal}

We define the angle, $\alpha_{\bx}$, between the local crystallographic axes and the axes of the ideal lattice~\cite{brock1992bond}:
$$\alpha_{\bx}=\frac{1}{2} \nabla \times \bu_{\bx} \,.$$
where we used the continuum representation.
In Fourier space we have:
\begin{equation*}
\hat \alpha_{\bq}=\frac{i}{2}( q_x \hat u_{\bq y}- q_y \hat u_{\bq x})
\end{equation*}
while the $\hat \alpha_{\bq}$ correlation reads
\begin{equation*}
\langle \hat \alpha_{\bq}  \hat \alpha_{-\bq}\rangle=  \frac{D\gamma}{4} \frac{q^2}{\omega_\bq^2}  \frac{1}{1+\frac{\tau}{\gamma}\omega_\bq^2}
\approx\frac{\sigma^2}{6 a^2} \frac{1}{1+ \ell^2 q^2 } \,,
\end{equation*}
where $\sigma^2=D\gamma/K$.
The real-space $\alpha_{\mathbf{x}}$-correlation function is given by
\begin{equation*}
\begin{aligned} 
 \langle ( \alpha_{\bx} - \alpha_{\bx'})^2   \rangle &\propto   \frac{1}{N^2}  \sum_{\bq}    e^{i\bq (\bx-\bx')}
 \frac{\sigma^2}{6 a^2} \frac{1}{1+\ell^2 q^2 } \\
&\propto \Bigl(\frac{\ell}{8\pi r} \Bigr)^{1/2} \frac {a^2} {\ell^2} e^{-  r/\ell} \,. 
\end{aligned}
 \end{equation*}
 where we have used the expansion for small $\mathbf{q}$ and the asymptotic behavior of $K_0(r/l)$.
Now, we consider the correlation function of $\psi_{6{\bx}}=e^{i6 \alpha_{\bx}}$ 
\begin{equation*}
\langle \psi_{6{\bx}} \psi^*_{6{\bx'}} \rangle=\langle e^{i6 \alpha_{\bx}} e^{-i6 \alpha_{\bx'}}   \rangle \,.
\end{equation*}
Using the form of the $\alpha$-correlation we find:
\begin{equation*}
\langle \psi_{6 \bx} \psi^*_{6 \bx'} \rangle= e^{-\frac{1}{2} \langle ( 6\alpha_{\bx} - 6\alpha_{\bx'})^2   \rangle} \,.
\end{equation*}
Consequently, the correlator of $\psi_6$ does not vanishes at infinity, i.e. the order is maintained since:
 \begin{equation*}
 \lim_{|\bx -\bx'|\to \infty}\langle \psi_{6 \bx} \psi^*_{ 6\bx'} \rangle=const \,.
 \end{equation*}
 This is the expected result since the harmonic lattice always maintains the sixfold coordination number
 and no disclinations can be created.
 We notice that for this model the velocity correlation and the $\alpha_{\bx}$-correlation have the same long-range behavior.

\bibliographystyle{apsrev4-1}

\bibliography{biblio.bib}

\end{document}